\documentclass[9pt,conference]{IEEEtran} 
\IEEEoverridecommandlockouts

\usepackage{cite}
\usepackage{amsmath,amssymb,amsfonts}
\usepackage{hyperref}
\usepackage{algorithmic}
\usepackage{graphicx}
\usepackage{textcomp}
\usepackage{xcolor}
\def\BibTeX{{\rm B\kern-.05em{\sc i\kern-.025em b}\kern-.08em
    T\kern-.1667em\lower.7ex\hbox{E}\kern-.125emX}}
\begin{document}

\title{Impairments are Clustered in Latents of Deep Neural Network-based Speech Quality Models\\
\thanks{The computations handling was enabled by the supercomputing resource Berzelius provided by National Supercomputer Centre at Link\"oping University and the Knut and Alice Wallenberg foundation.}
}









\author{
    \IEEEauthorblockN{Fredrik Cumlin \IEEEauthorrefmark{1}, Xinyu Liang \IEEEauthorrefmark{2}, Victor Ungureanu
    \IEEEauthorrefmark{3}, Chandan K. A. Reddy \IEEEauthorrefmark{3}, Christian Sch\"uldt \IEEEauthorrefmark{3}, Saikat Chatterjee \IEEEauthorrefmark{1}} 

    \IEEEauthorblockA{\IEEEauthorrefmark{1} School of Electrical Engineering and Computer Science, KTH Royal Institute of Technology, Sweden}
    \IEEEauthorblockA{\IEEEauthorrefmark{2} HCLTech, Sweden \,\,\, \IEEEauthorrefmark{3} Google LLC}
    \IEEEauthorblockA{\IEEEauthorrefmark{1}  \{fcumlin, sach\}@kth.se, 
    \IEEEauthorrefmark{2} hopeliang@icloud.com, \IEEEauthorrefmark{3} \{ungureanu, chandanka, cschuldt\}@google.com}
}

\maketitle

\begin{abstract}
In this article, we provide an experimental observation: Deep neural network (DNN) based speech quality assessment (SQA) models have inherent latent representations where many types of impairments are clustered. While DNN-based SQA models are not trained for impairment classification, our experiments show good impairment classification results in an appropriate SQA latent representation. We investigate the clustering of impairments using various kinds of audio degradations that include different types of noises, waveform clipping, gain transition, pitch shift, compression, reverberation, etc. To visualize the clusters we perform classification of impairments in the SQA-latent representation domain using a standard k-nearest neighbor (kNN) classifier. We also develop a new DNN-based SQA model, named DNSMOS+, to examine whether an improvement in SQA leads to an improvement in impairment classification. The classification accuracy is $94\%$ for LibriAugmented dataset with 16 types of impairments and $54\%$ for ESC-50 dataset with $50$ types of real noises.
\end{abstract}


\section{Introduction}
\label{sec:intro}

Auditory-motivated perceptual features are extensively used in audio signal processing. Examples of time-tested features are Mel-frequency cepstral coefficients (MFCCs) \cite{MFCC}, perceptual linear prediction (PLP) \cite{PLP}, and Gammatone frequency cepstral coefficients (GFCCs) \cite{GFCC}. The use of MFCC features is a standard in speech recognition, speaker identification, and many other speech processing tasks \cite{MFCC}.

Interesting experimental evidence exists that noise sounds, such as coughing, train noise, mouse-clicking, etc., get clustered in the MFCC feature vector domain \cite{esc50}. To visualize the clusters, a baseline kNN classifier was used in the MFCC feature space, to correctly classify noise types $66.7\%$ in ESC-10 and $32.2\%$ in ESC-50. Note that the signal processing steps to generate MFCC feature vectors are auditory-motivated, and not explicitly designed for noise classification. Therefore, the experimental evidence suggests that noise classification happens in an auditory-motivated feature domain (the MFCC domain). 

Speech quality assessment (SQA) is the task of estimating speech quality on a scale based on human perception. Speech quality is affected by different impairments, such as varying types of noises, reverberation, codec artifacts, etc. 
Currently, the best-performing SQA models are designed using deep neural networks (DNNs) \cite{LDNet, LaMOSNet, DeePMOS, NISQA, UTMOS, ZevoMOS}.

In this article, we show a phenomenon: DNN-based SQA models produce internal (latent) feature vectors where many types of impairments are clustered. Note that SQA models are not engineered for clustering impairments. The training and learning algorithms for the DNN-based SQA methods do not use any data and/or optimization problems related to impairment clustering. The clustering of impairments turns out to be an inherent by-product. To the best of the authors' knowledge, our work is the first experimental evidence of the phenomenon. Our contributions are as follows:
\begin{enumerate}
    \item To visualize the clustering of impairments in a suitable latent feature space, we perform a classification task to identify the types of impairments. For this, we employ a standard k-Nearest Neighbors (kNN) method, similar to the approach used by \cite{esc50}. Our experiments demonstrate that the impairment classification accuracy is significantly higher in the latent domain of DNN-based SQA models compared to the traditional MFCC domain. For instance, in the ESC-50 dataset, which contains $50$ distinct noise types, we achieve a classification accuracy of $27\%$ in the latent domain of a DNN-based SQA model, compared to just $18\%$ in the MFCC domain.

    \item We hypothesize that an improvement in SQA performance correlates with an enhancement in impairment classification accuracy. To test this hypothesis, we first evaluate impairment classification accuracy in the latent domain of an existing DNN-based SQA model, DNSMOS \cite{DNSMOS}. We then apply an improved version, DNSMOS Pro \cite{DNSMOSp}, and examine its impact on impairment classification. To adapt DNSMOS Pro for a regression task (instead of the probabilistic approach proposed in the original paper), we introduce slight modifications and refer to the resulting model as DNSMOS+. Using the LibriAugmented dataset, we observe that DNSMOS achieves an SQA performance of $0.89$, as measured by the Pearson correlation coefficient (PCC), with an impairment classification accuracy of $86\%$. The enhanced DNSMOS+ model improves SQA performance to $0.93$ (PCC), with a corresponding increase in impairment classification accuracy to $94\%$, thereby supporting our hypothesis.
\end{enumerate}


\subsection{DNN-based Speech Quality Assessment Models}

In this subsection, we provide a brief literature review on SQA models, primarily on non-intrusive DNN-based SQA models.

For a non-intrusive SQA, the goal is to predict the speech quality given only a degraded speech signal (i.e., in the absence of a reference signal). The SQA models also can be intrusive, for example, PESQ \cite{PESQ}, ViSQOL \cite{visqol}.

Over the past years, DNN-based non-intrusive SQA models have become dominant. Early works include AutoMOS \cite{AutoMOS} and QualityNet \cite{QualityNet}, where the first is trained using mean-opinion-score (MOS) \cite{p808} and the latter is trained using PESQ \cite{PESQ} labels. Later works have been trained end-to-end with MOS labels, such as DNSMOS, or utilizing individual scores for increased performance \cite{LDNet, DeePMOS, LaMOSNet}. Further, there are unsupervised feature extraction-based SQA models \cite{w2v_mos, moosenet, UTMOS, LE-SSL-MOS}.



Some methods have explored measuring the quality from more than one perspective, such as DNSMOSp835 
\cite{dnsmos_p835} and NISQA \cite{NISQA}. DNSMOSp835 is trained on scores collected under ITU-T Rec. P.835 \cite{p835}, which aims to emulate the subjective evaluation for the amount of noise and the impairment of the speech together with the overall speech quality. NISQA hypothesizes that overall speech quality can be described by four different quality 'dimensions': noisiness, coloration, discontinuity, and loudness. However, none of the mentioned SQA models provides the specific \emph{cause} of poor speech quality, instead, only scalar estimates of the quality are provided.

In this article, we use the DNSMOS models \cite{DNSMOS, dnsmos_p835} to study how impairments are represented in latent space. The choice of DNSMOS is due to its low complexity. Then we also study how the impairments are represented in the slightly improved model DNSMOS Pro \cite{DNSMOSp}.

\section{Method}
\label{sec:method}

\begin{figure}[t]

\begin{minipage}[b]{.49\linewidth}
  \centering
  \centerline{\includegraphics[width=4.1cm]{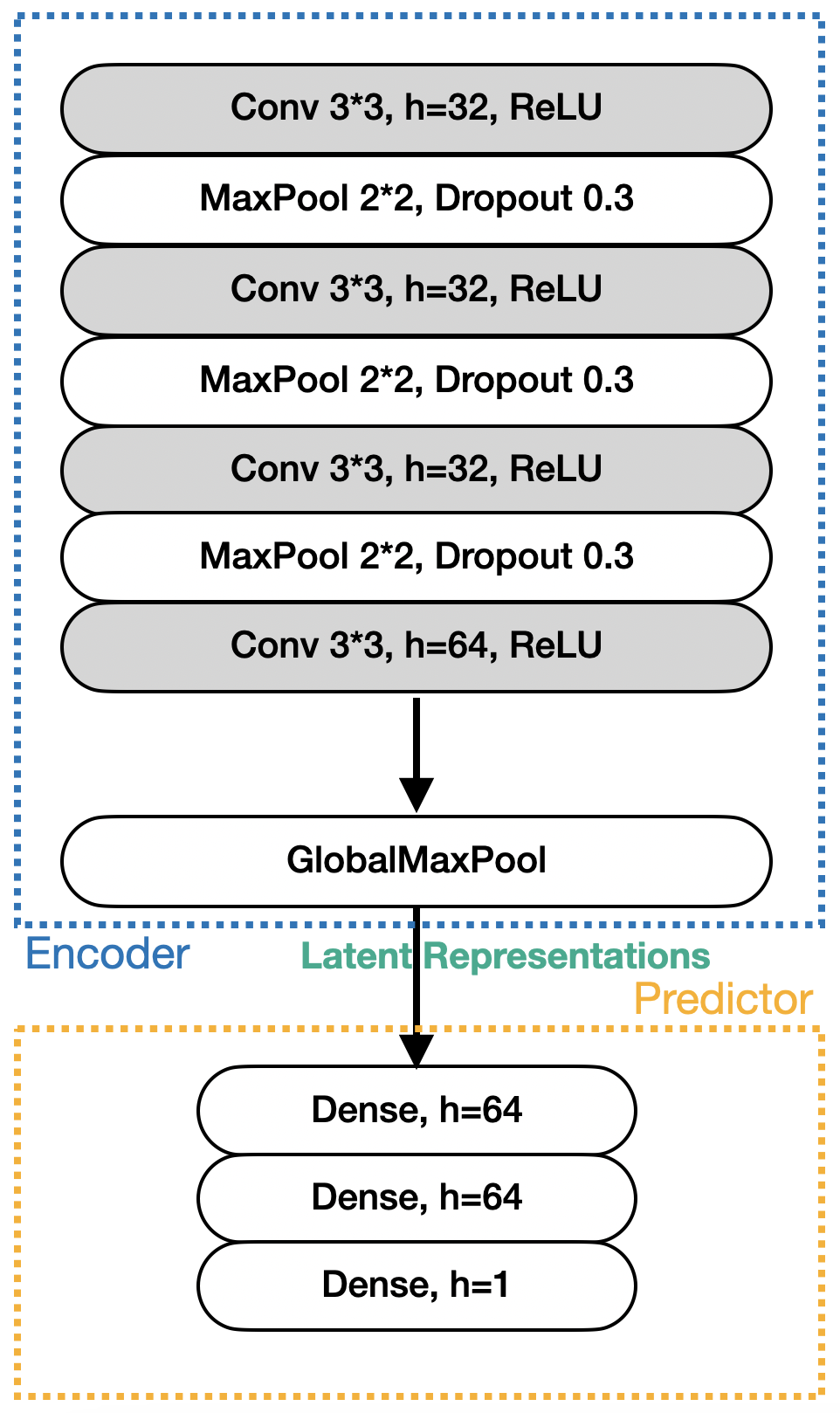}}
  \centerline{(a) DNSMOS architecture.}\medskip
\end{minipage}
\hfill
\begin{minipage}[b]{.49\linewidth}
  \centering
  \centerline{\includegraphics[width=4.1cm]{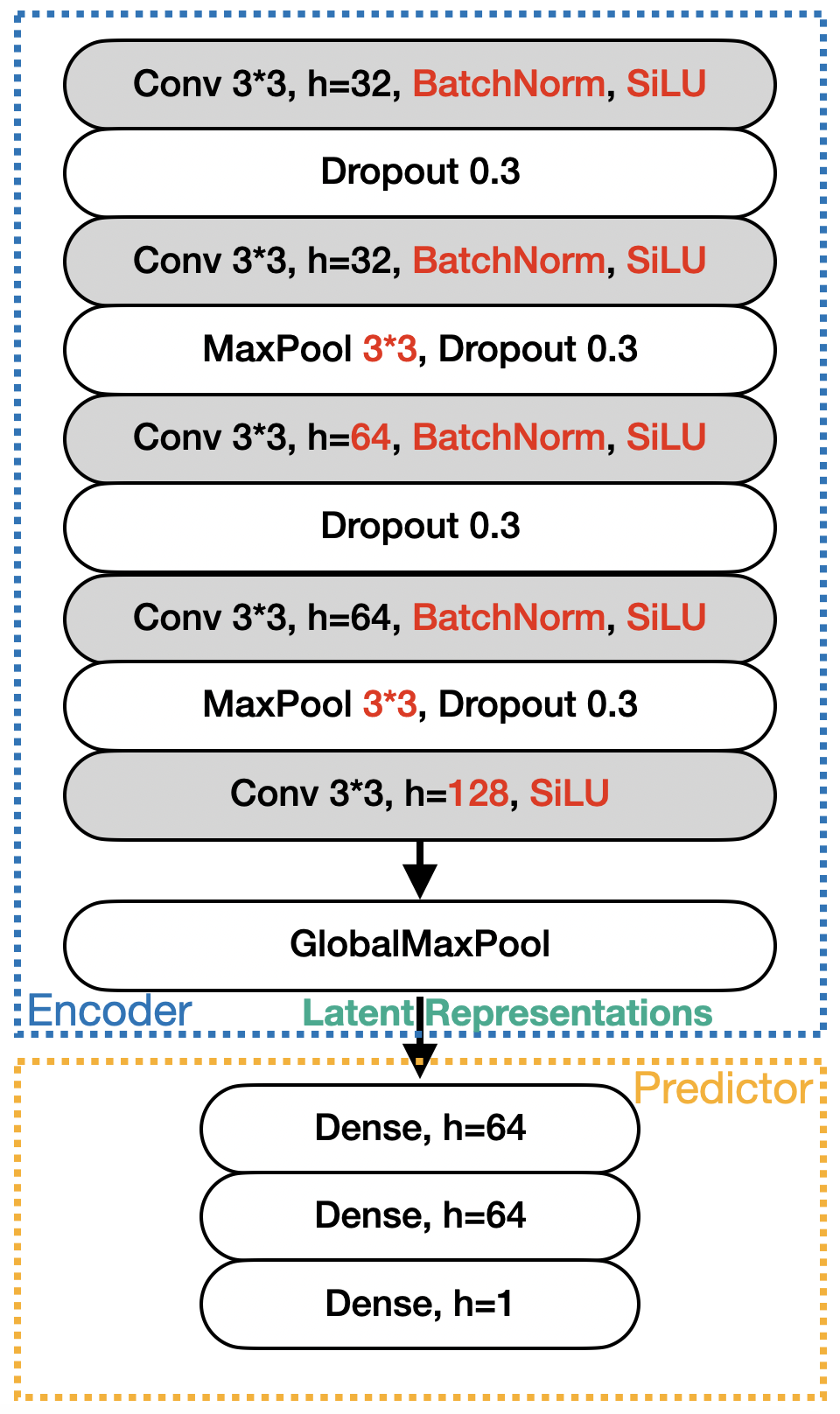}}
  \centerline{(b) DNSMOS+ architecture.}\medskip
\end{minipage}
\caption{Architectures of DNSMOS and DNSMOS+.}
\label{fig:arch}
\end{figure}

We start with the basic DNSMOS \cite{DNSMOS} model, which is a popular low-complexity non-intrusive SQA method. It uses the Mel spectrogram as input and outputs a scalar value: the speech quality in terms of MOS. The DNSMOS architecture consists of two parts, an encoder, and a predictor, as shown in Figure \ref{fig:arch} (a). The encoder consists mainly of convolutional and pooling layers and produces latent representations of the input spectrogram. The predictor consists of dense layers and maps the latent representation to a scalar MOS value. The training of DNSMOS is an end-to-end supervised learning mechanism where labeled data is used. The labeled data is comprised of a pairwise input spectrogram and the corresponding MOS output as a label.  

\subsection{Clustering of impairments in latent representation}


A DNSMOS predictor provides SQA in terms of MOS using the latent representation. If a speech signal has no impairments then the MOS output is expected to be good. On the other hand, the presence of impairments is expected to provide a poor MOS value. That means the latent representation has some information about the impairments. Then natural questions arise: 
\begin{enumerate}
    \item How is the information about the impairments reflected in the latent representation space? 
    \item Can we visualize the effect of impairments in the latent representation space?
    \item Do different kinds of impairments form clusters in the latent space? If so, are the clusters distinct or highly overlapping?
\end{enumerate}

The dimension of the latent representation vector is much smaller than the input spectrogram dimension. In our case, we use latent dimension as either $64$ or $128$. Therefore, visualization of the latent representation of a speech signal under an impairment or multiple types of impairments is not trivial. 

An indirect approach to visualize the latent representation and forming of clusters is via a classification study: we create a labeled dataset of speech using different types of impairments and measure the accuracy of a kNN classifier. 


\subsection{Dataset construction}\label{subsec:libriaugmented}

For a more controlled experimental setup, we also generate our pseudo-labeled dataset and train a non-intrusive SQA model end-to-end from speech clip features to speech quality. The dataset is given by $\mathcal{D}={(\mathbf{x}, y, z)}_{\mathbf{x}\in \mathcal{X}}$, where $\mathbf{x}$ is an impaired signal, $y$ is speech quality label from an intrusive quality measure, and $z$ is the impairment class applied to a clean speech signal to produce $x$.

We use the LibriSpeech \cite{librispeech} dataset as the source of clean speech data, and introduce different types of impairments to the speech clips. This clean speech dataset consists of $100$ hours of speech, a total of $28\,539$ speech clips, with an average duration of $13$ s and sample rate $16$ kHz. We crop or repetitively pad all clean speech clips to $10$ s.

The impairments are added through augmentations supported by the Audiomentations library\footnote{https://github.com/iver56/audiomentations}. 
We corrupt the clean signals with $9$ different single impairments and 6 double impairments to make our simulated dataset closer to real-world scenarios. The chosen ratio of impairments as well as the parameters, are listed in Table \ref{table:dataset}. For the additive background noise, we randomly sample it from Demand\footnote{https://zenodo.org/records/1227121} and FreeSound\footnote{https://freesound.org}, which in total consists of $23$ different noise data sources. We call the constructed dataset \textbf{LibriAugmented}, consisting of $57\,078$ speech clips and $16$ impairment classes (considering the identity mapping as a class). The subset of LibriAugmented in which speech clips are corrupted by at most one impairment is noted as LibriAugmented$_{one}$. The complete LibriAugmented contains each clean speech clip from LibriSpeech exactly twice.


\begin{table}[ht!]
\caption{The LibriAugmented dataset.}
\begin{center}
\begin{tabular}{cc}
\hline
     \textbf{Ratio} & \textbf{Impairment (Parameters)} \\ \hline 
     $0.1$ & Identity \\
     $0.1$ & AddBackgroundNoise (snr=$-10\sim 15$ dB) \\
     $0.1$ & Clippingimpairment (percentile=$10\sim40\%$) \\
     $0.1$ & GainTransition (gain=$-60\sim20$ dB) \\
     $0.1$ & LowPassFilter (cutoff\_freq=$0.5\sim1$ kHz) \\
     $0.1$ & Mp3Compression (bit\_rate=$8\sim14$) \\
     $0.1$ & PitchShift (semitones=$-4\sim4$ kHz) \\
     $0.1$ & RoomSimulator (rt60=$0.8\sim1.5$ s) \\
     $0.1$ & TimeMask (band\_part=$0.2\sim0.5$) \\
     $0.1$ & TimeStretch (rate=$0.5\sim2$) \\
     \hline
     $0.167$ & AddBackgroundNoise + RoomSimulator \\
     $0.167$ & AddBackgroundNoise + LowPassFilter \\
     $0.167$ & AddBackgroundNoise + TimeStretch \\
     $0.167$ & RoomSimulator + Mp3Compression \\
     $0.167$ & PitchShift + LowPassFilter \\
     $0.167$ & GainTransition + TimeMask \\ \hline
\end{tabular}
\label{table:dataset}
\end{center}
\end{table}

We label the dataset using PESQ \cite{PESQ} and ViSQOL v3 \cite{visqol}, both of which are designed to simulate MOS measures. We split the data into three sets: a training dataset, a validation dataset, and a testing dataset. We explicitly make the split between one-impairment and two-impairment subsets separately, so that each half contains $22\, 539$, $3\, 000$, and $3\, 000$ speech clips in train/val/test respectively. The clean speech utterances are disjoint in the train/val/test split, meaning that for example if we have a speech clip in the one-impairment train set, then the same clean speech (but distorted with two impairments) is also in the two-impairment train set.

\subsection{DNSMOS+ design}

We use an improved DNSMOS model, namely DNSMOS Pro \cite{DNSMOSp}, to study how improvements in quality assessment relate to the clustering of impairments. We change the model slightly from the probabilistic definition given in \cite{DNSMOSp}, and we call this model DNSMOS+. Changes of DNSMOS+ architecture from the DNSMOS architecture include (1) STFT preprocessing instead of Mel spectrogram as input, (2) an extra convolutional layer, (3) SiLU (Swish) activation function instead of ReLU, (4) more sparse max-pooling layers with larger kernels, and (5) the addition of batch normalization layers. A comparison of the architectures is illustrated in Figure \ref{fig:arch}.


\section{Experiments}

\begin{table*}
\caption{Performance results on LibriAugmented and ESC-50. All self-implemented algorithms are run $10$ times with different seeds, and the mean and standard deviation are reported. When trained towards PESQ labels, we also evaluate on PESQ labels on the LibriAugmented dataset.}
\begin{center}
\resizebox{\textwidth}{!}{%
\begin{tabular}{c|c|c|cccc|cc}
\hline
     & & & \multicolumn{4}{c|}{LibriAugmented} & \multicolumn{2}{c}{ESC-50} \\
    {Model} & {Labels} & {Training data} & {MSE} & {PCC} & {SRCC} & {Top-1 Acc} & {Top-1 Acc} & {Top-3 Acc} \\ \hline
    {Random projection} & - & - & - & - & - & {$0.08{\scriptstyle\pm 0.01}$} & {$0.03{\scriptstyle\pm 0.01}$} & {$0.07{\scriptstyle\pm 0.01}$} \\
    {MFCC} & - & - & - & - & - & {$0.30$} & {$0.18$} & {$0.36$} \\
    {DNSMOS}  & {MOS} & {DNS data \cite{DNSMOS}} & {$1.13$} & {$0.41$} & {$0.43$} & {$0.59$} & {$0.27$} & {$0.52$} \\
    {DNSMOSp835}  & {MOS} & {DNS data \cite{dnsmos_p835}} & {$0.80$} & {$0.59$} & {$0.64$} & {$0.74$} & {$0.27$} & {$0.53$} \\
    {DNSMOS}  & {ViSQOL} & {LibriAugmented} & {$0.28{\scriptstyle\pm 0.01}$} & {$0.89{\scriptstyle\pm 0.01}$} & {$0.90{\scriptstyle\pm 0.01}$} & {$0.86{\scriptstyle\pm 0.01}$} & {$0.42{\scriptstyle\pm 0.02}$} & {$0.67{\scriptstyle\pm 0.01}$} \\
    {DNSMOS+}  & {ViSQOL} & {LibriAugmented$_{one}$} & {$0.21{\scriptstyle\pm 0.04}$} & {$0.90{\scriptstyle\pm 0.01}$} & {$0.89{\scriptstyle\pm 0.02}$} & {$0.93{\scriptstyle\pm 0.01}$} & {$0.51{\scriptstyle\pm 0.02}$} & {$0.78{\scriptstyle\pm 0.02}$} \\
    {DNSMOS+}  & {PESQ} & {LibriAugmented} & {$0.33{\scriptstyle\pm 0.15}$} & {$0.87{\scriptstyle\pm 0.05}$} & {$0.87{\scriptstyle\pm 0.01}$} & {$0.93{\scriptstyle\pm 0.01}$} & {$0.49{\scriptstyle\pm 0.02}$} & {$0.76{\scriptstyle\pm 0.01}$} \\
    {DNSMOS+}  & {ViSQOL} & {LibriAugmented} & {$\mathbf{0.19}{\scriptstyle\pm 0.03}$} & {$\mathbf{0.93}{\scriptstyle\pm 0.01}$} & {$\mathbf{0.94}{\scriptstyle\pm 0.01}$} & {$\mathbf{0.94}{\scriptstyle\pm 0.01}$} & {$\mathbf{0.54}{\scriptstyle\pm 0.01}$} & {$\mathbf{0.80}{\scriptstyle\pm 0.01}$} \\ \hline
\end{tabular}}\label{table:performance}
\end{center}
\end{table*}

\subsection{Datasets}

In our experiments, we use two different datasets: the LibriAugmented dataset as described in \ref{subsec:libriaugmented}, and the Environmental Sound Classification 50 (ESC-50) dataset \cite{esc50}. LibriAugmented is used as a training dataset for the non-intrusive models, and to study the clustering of different impairments. ESC-50 consists of environmental noise files and is used to study the clustering of noises in the latent space of a non-intrusive SQA model.

The ESC-50 dataset consists of $50$ different noise categories with $40$ samples in each category, resulting in a total of $2\, 000$ noise clips. All signals are $5$ s long and sampled at $16$ kHz. Examples include noise types are mouse-clicking, cow, thunderstorm, laughing, vacuum cleaner, baby crying, coughing, and helicopter. Human accuracy on this dataset is $\sim 81.3\%$ \cite{esc50}.



\subsection{Feature extraction}
The feature extraction of the speech clips, before processing by the DNN, is similar to DNSMOS \cite{DNSMOS}. The inputs are log-magnitude spectrograms with a window duration of $20$ ms and a hop duration of $10$ ms, using a Hann window, and at a sample rate of $16$ kHz. Furthermore, the values in the spectrogram are clipped to the interval $[-7, 7]$. This is to avoid having large values making training unstable.

\subsection{Experimental setup}

We design four training experiments: (1) DNSMOS (architecture) on LibriAugmented dataset using ViSQOL labels; (2) DNSMOS+ on LibriAugmented$_{one}$ using ViSQOL labels; (3) DNSMOS+ on LibriAugmented using PESQ labels; (4) DNSMOS+ on LibriAugmented using ViSQOL labels.

The experiments were designed from an observational point of view, where the latent clustering is studied in these configurations\footnote{Code can be found at \url{https://github.com/fcumlin/sqa\_latent\_classification}.}.

For all experiments, we train end-to-end from speech clip features to quality labels. We use the mean-square-error (MSE) as a loss function. An Adam optimizer with $\beta_1=0.9$ and $\beta_2=0.999$ \cite{Adam} was used, together with a learning rate of $10^{-4}$ and batch size of $32$, as per DNSMOS \cite{DNSMOS}. We train for $500$ epochs and extract the model for analysis once it has been fully trained. One epoch means one iteration over the whole training dataset. Training is done on one Nvidia A100 40GB GPU card, and training one model takes $\sim 30$ hours to complete. Testing is done on the LibriAugmented test data.

\subsubsection{Performance on predicting quality label}
As performance measures we report the mean squared error (MSE), Pearson Correlation Coefficient (PCC), and Spearman Rank Correlation Coefficient (SRCC). These measurements are used to compare the predictions with the ViSQOL labels on the test data. If the model is trained on PESQ labels, we use the PESQ labels on the test data instead.

\subsubsection{Performance on predicting impairments}
To measure the latent clustering of the impairments we do as explained in the method Section~\ref{sec:method}. Given the LibriAugmented test dataset, we have impaired speech clips and impairment class label pairs. Some speech clips have been distorted by two impairments, and we consider each such pair as its own class. This means each speech clip has exactly one impairment label.

We further split this LibriAugmented test set for training and testing a kNN model, and this split is done in a stratified way so that the proportion of different classes in each split is approximately the same. We use 70$\%$ of data in training and 30$\%$ in test. A kNN, using $k=15$, is trained on the latent representations of the training subset. We measure the accuracy of predicting the augmentation on the test subset.

\subsubsection{Performance on predicting noise class}
To measure the classification performance of different noises we use the ESC-50 dataset. We partition this dataset into two sets, train and test, in a similar stratified partitioning procedure as described in the previous paragraph. A kNN, using $k=15$, is also trained similarly, and we measure the accuracy of predicting the noise class on the test subset.

\subsubsection{Baseline and models from the literature}

For the classification task of predicting the impairment class in the LibriAugmented dataset and the noise class in the ESC-50 dataset, we use three additional methods for comparison.

As a baseline, we consider random projection \cite{random_projection}. This is done by initializing a matrix $A$ in $\mathbb{R}^{d\times 128}$, where $d$ is the number of samples of an audio clip. The initialization is done so the columns in the matrix are realizations of independent and identically distributed normal vectors with an expected $l_2$ length of $1$. The mapping of an audio clip $x$ to a $128$-dimensional vector is given by $x\mapsto Ax$. Subsequently, classification analysis is done using the image of a speech clip under the random projection as input to the kNN model. 

We also include classification results when using MFCC as an input feature. This is done by computing the MFCCs of the speech clips using $12$ bins and using the image as input to the explained kNN procedure.

We furthermore study pre-trained models in the literature, namely DNSMOS \cite{DNSMOS} and DNSMOSp835 \cite{dnsmos_p835}\footnote{The models can be found in \url{https://github.com/microsoft/DNS-Challenge/tree/master/DNSMOS}.}. The models were extracted as-is; no additional training of the model weights was done. The latent representations after the GlobalMaxPool layer were used as input to train a kNN as explained.

\subsection{Results}

\begin{figure}[t]
  \centering
  \includegraphics[width=1\linewidth]{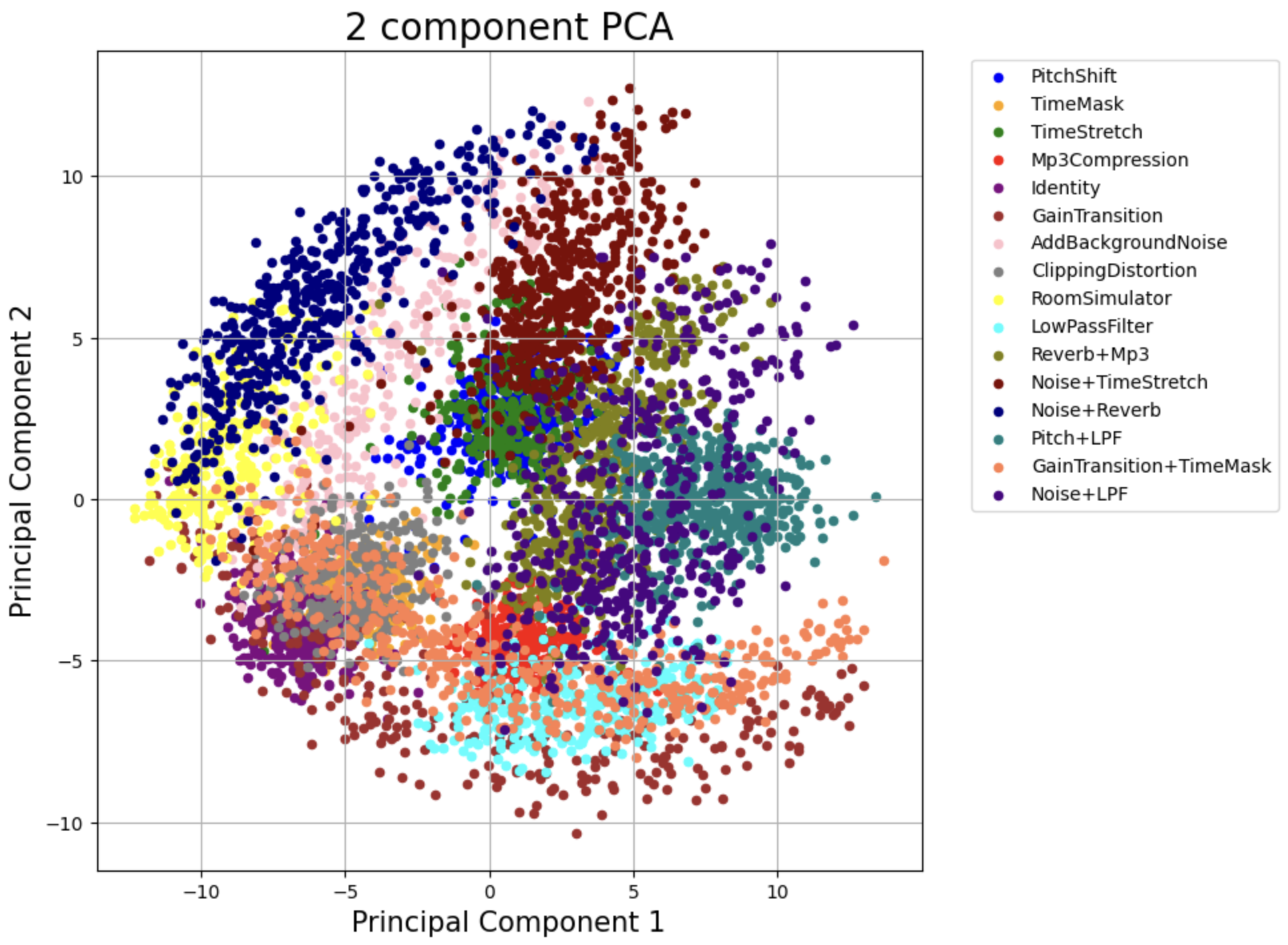}
  \caption{PCA visualization, using output dimension of $2$, of latent features of the LibriAugmented datasets using DNSMOS+ (ViSQOL labels).}
  \label{fig:pca2}
\end{figure}

The results of speech quality prediction performance and classification performance are shown in Table \ref{table:performance}. Note that the main interest is the observation of the clustering of impairments, not speech quality prediction performance per se. In Fig. \ref{fig:pca2}, a 2-component PCA of the latent representation of LibriAugmented dataset in the DNSMOS+ model is shown.

We notice that models trained on LibriAugmented and models from literature outperform random projection in classifying the impairment and noise class respectively. This gives evidence of an inductive bias (``tendency'') of the DNNs to cluster different impairments.

Furthermore, we can see a clear difference in the noise classification accuracy when comparing the DNSMOS models from the literature to the LibriAugmented trained ones. One reason for this can be about the data: The DNS dataset consists of $600$ noisy speech clips processed by $>200$ noise suppressors and $40$ noise suppressors for DNSMOS and DNSMOSp835 respectively. Our training dataset consists of $22\, 539$ unique speech clips processed with one or two impairments, which might increase the clustering of noises. Another plausible explanation is that the DNS data used has been processed by noise suppressors, and so could suppress the noises to a degree where clustering of noises is exhibited to a smaller extent. The reason herein is an interesting topic for future study.

Considering the trained models, it shows that DNSMOS+ is a more suitable architecture than DNSMOS. One reason could be the increased dimension of the latent space ($64$-dimensional vector for DNSMOS to $128$-dimensional vector for DNSMOS+). This potentially increases the expressivity for DNSMOS+. Suitably applying normalization and other changes makes DNSMOS+ have higher quality prediction measures, and also higher classification accuracy.

We also notice that both ViSQOL and PESQ labels demonstrate their suitability for being used as labels in the SQA task and the classification task (the last two rows in Table \ref{table:performance}). However, ViSQOL labels seem more suitable compared to PESQ labels, as the average performance is higher with a lower standard deviation across runs when using ViSQOL labels compared to PESQ labels. The reason could be that ViSQOL(v3) is a better MOS emulator compared to PESQ for the considered impairments and datasets, which is consistent with the authors' subjective opinions when listening to the created dataset; ViSQOL is more aligned with our judgment compared to PESQ. 

In Fig. \ref{fig:pca2}, a 2-component PCA visualization of the latent representations from the LibriAugmented dataset using the DNSMOS+ model is presented. The labels correspond to the distortions applied to the signals before processing. The plot reveals that speech signals with similar distortions tend to cluster together. Importantly, this clustering is not driven by the quality labels, as the quality distributions are roughly equal across different distortion classes. Therefore, the model's ability to cluster these distortions appears to be an emergent behavior, rather than a consequence of any single distortion consistently producing higher or lower quality signals.




\section{Conclusion}

In this paper, we provided an experimental observation that impairments are clustered in the latent space of DNN-based non-intrusive speech quality models. These models are trained end-to-end to predict only speech quality scores. Using DNSMOS from the literature, we showed $~30\%$ top-1 noise accuracy and $~60-75\%$ top-1 impairment accuracy. Using a regression-suitable DNSMOS Pro architecture and training on a more diverse dataset, we achieved $54\%$ top-1 noise accuracy and $94\%$ top-1 impairment accuracy. Important to note is that \emph{the models were not trained for impairment classification} - they were only trained to predict quality. The results, together with the low accuracy when using random linear projection, support the hypothesis that DNN-based SQA models have inherent latent representations where many types of impairments are clustered.


\begin{thebibliography}{00}
\bibitem{MFCC} Z. K. Abdul and A. K. Al-Talabani, ``Mel Frequency Cepstral Coefficient and its Applications: A Review,'' \textit{IEEE Access}, vol. 10, pp. 122136-122158, 2022, doi: 10.1109/ACCESS.2022.3223444.
\bibitem{PLP} F. H{\"o}nig, G. Stemmer, C. Hacker, and F. Brugnara, ``Revising Perceptual Linear Prediction (PLP),'' in \textit{Proc. Interspeech}, 2005.
\bibitem{GFCC} B. Ayoub, J. Kharroubi, and A. Zarghili, ``Gammatone frequency cepstral coefficients for speaker identification over VoIP networks,'' in \textit{Proc. 2016 Int. Conf. Information Technology for Organizations Development (IT4OD)}, 2016, pp. 1-5, doi: 10.1109/IT4OD.2016.7479293.
\bibitem{esc50} K. J. Piczak, ``ESC: Dataset for Environmental Sound Classification,'' in \textit{Proc. 23rd Annual ACM Conf. Multimedia}, Brisbane, Australia, Oct. 2015, pp. 1015--1018, doi: 10.1145/2733373.2806390.
\bibitem{LaMOSNet} F. Cumlin, C. Sch\"uldt, and S. Chatterjee, ``Latent-based Neural Net for Non-intrusive Speech Quality Assessment,'' in \textit{Proc. 2023 33rd European Signal Processing Conference (EUSIPCO)}, Sept. 2023, pp. 36--40.
\bibitem{DNSMOS} C. Reddy, V. Gopal, and R. Cutler, ``DNSMOS: A Non-Intrusive Perceptual Objective Speech Quality Metric to Evaluate Noise Suppressors,'' in \textit{Proc. IEEE Int. Conf. Acoust., Speech, Signal Process. (ICASSP)}, Toronto, ON, Canada, June 6-11, 2021, doi: 10.1109/ICASSP39728.2021.9414878.
\bibitem{DNSMOSp} F. Cumlin, X. Liang, V. Ungureanu, C. Reddy, C. Sch\"uldt, S.  Chatterjee, ``DNSMOS Pro: A Reduced Size DNN for Probabilistic MOS of Speech,'' in \textit{Proc. Interspeech 2024, 25th Annu. Conf. Int. Speech Commun. Assoc.}, Kos Island, Greece, Sept. 1-5, 2024.
\bibitem{LDNet} W.-C. Huang, E. Cooper, J. Yamagishi, and T. Toda, ``LDNet: Unified Listener Dependent Modeling in MOS Prediction for Synthetic Speech,'' in \textit{Proc. IEEE Int. Conf. Acoust., Speech, Signal Process. (ICASSP)}, 2022, doi: 10.1109/ICASSP43922.2022.9747222.
\bibitem{NISQA} G. Mittag, B. Naderi, A. Chehadi, and S. Möller, ``NISQA: A Deep CNN-Self-Attention Model for Multidimensional Speech Quality Prediction with Crowdsourced Datasets,'' in \textit{Proc. Interspeech 2021}, Aug. 2021, doi: 10.21437/Interspeech.2021-299.
\bibitem{QualityNet} S.-W. Fu, Y. Tsao, H.-T. Hwang, and H.-M. Wang, ``Quality-Net: An End-to-End Non-intrusive Speech Quality Assessment Model Based on BLSTM,'' in \textit{Proc. Interspeech 2018, 19th Annu. Conf. Int. Speech Commun. Assoc.}, Hyderabad, India, Sept. 2-6, 2018, doi: 10.21437/Interspeech.2018-1802.
\bibitem{DeePMOS} X. Liang, F. Cumlin, C. Sch\"uldt, and S. Chatterjee, ``DeePMOS: Deep Posterior Mean-Opinion-Score of Speech,'' in \textit{Proc. Interspeech 2023}, pp. 526--530.
\bibitem{AutoMOS} B. Patton, Y. Agiomyrgiannakis, M. Terry, K. Wilson, R. A. Saurous, and D. Sculley, ``AutoMOS: Learning a Non-Intrusive Assessor of Naturalness-of-Speech,'' in \textit{Proc. NIPS 2016 End-to-End Learning for Speech and Audio Processing Workshop}, 2016.
\bibitem{UTMOS} T. Saeki, D. Xin, W. Nakata, T. Koriyama, S. Takamichi, and H. Saruwatari, ``UTMOS: UTokyo-SARULAB System for VoiceMOS Challenge 2022,'' \textit{arXiv preprint arXiv:2204.02152}, 2022.
\bibitem{Adam} D. P. Kingma and J. Ba, ``Adam: A Method for Stochastic Optimization,'' \textit{arXiv preprint arXiv:1412.6980}, 2014.
\bibitem{LE-SSL-MOS} Z. Qi, X. Hu, W. Zhou, S. Li, H. Wu, J. Lu, and X. Xu, ``LE-SSL-MOS: Self-Supervised Learning MOS Prediction with Listener Enhancement,'' in \textit{Proc. 2023 IEEE Int. Conf. Multimedia and Expo (ICME)}, 2023.
\bibitem{ZevoMOS} A. Stan, ``The Zevomos Entry to VoiceMOS Challenge 2022,'' \textit{arXiv preprint arXiv:2206.07448}, 2022.
\bibitem{PESQ} A. W. Rix, J. G. Beerends, M. P. Hollier, and A. P. Hekstra, ``Perceptual evaluation of speech quality (PESQ) - a new method for speech quality assessment of telephone networks and codecs,'' in \textit{Proc. ICASSP 2001}, vol. 2, 2001, doi: 10.1109/ICASSP.2001.941023.
\bibitem{visqol} M. Chinen, F. S. C. Lim, J. Skoglund, N. Gureev, F. O'Gorman, and A. Hines, ``ViSQOL v3: An open source production ready objective speech and audio metric,'' arXiv, 2020, doi: 10.48550/ARXIV.2004.09584.
\bibitem{p835} ITU-T Recommendation P.835, ``Subjective test methodology for evaluating speech communication systems that include noise suppression algorithm,'' International Telecommunication Union, Geneva, 2003.
\bibitem{p808} ITU-T Recommendation P.808, ``Subjective evaluation of speech quality with a crowdsourcing approach,'' International Telecommunication Union, Geneva, 2018.
\bibitem{moosenet} O. Platek and O. Dusek, ``MooseNet: A Trainable Metric for Synthesized Speech with a {PLDA} Module,'' in \textit{Proc. 12th Speech Synthesis Workshop (SSW)}, 2023.
\bibitem{w2v_mos} E. Cooper, W.-C. Huang, T. Toda, and J. Yamagishi, ``Generalization Ability of MOS Prediction Networks,'' in \textit{Proc. ICASSP 2022}, 2022, pp. 8442--8446, doi: 10.1109/ICASSP43922.2022.9746395.
\bibitem{dnsmos_p835} C. K. A. Reddy, V. Gopal, and R. Cutler, ``Dnsmos P.835: A non-intrusive perceptual objective speech quality metric to evaluate noise suppressors,'' in \textit{Proc. ICASSP 2022}, 2022, pp. 886--890.
\bibitem{librispeech} V. Panayotov, G. Chen, D. Povey, and S. Khudanpur, ``Librispeech: An ASR corpus based on public domain audio books,'' in \textit{Proc. ICASSP 2015}, 2015, doi: 10.1109/ICASSP.2015.7178964.
\bibitem{random_projection} I. Kononenko and M. Kukar, ``Data preprocessing,'' in \textit{Machine Learning and Data Mining}, I. Kononenko and M. Kukar, Eds. Woodhead Publishing, 2007, pp. 181-211, doi: 10.1533/9780857099440.181.






\end{thebibliography}
\end{document}